\documentclass[trackchanges]{aastex701}

\usepackage{amsmath}

\usepackage{etoolbox}
\usepackage[english]{babel}
\usepackage[autostyle, english = american]{csquotes}
\MakeOuterQuote{"}

\journalinfo{\today}
\makeatletter
\nobreakafterkeywordsfalse
\def\ps@plain{%
  \let\@oddhead\@empty
  \let\@evenhead\@empty
  \def\@oddfoot{\hfill\thepage}%
  \let\@evenfoot\@oddfoot
}
\def\ps@myheadings{%
  \let\@oddhead\@empty
  \let\@evenhead\@empty
  \def\@oddfoot{\hfill\thepage}%
  \let\@evenfoot\@oddfoot
}
\makeatother
\begin{document}
\pagenumbering{gobble}

\fontsize{11}{16.5}\selectfont

\title{\Large Preparing for the Next Carrington: Spatiotemporal Agent-Based Modeling for Safeguarding Satellite Infrastructure Under Extreme Space Weather Disturbances\\[1.5em]}

\author[orcid=0009-0005-5282-0013 ]{Rushil Kukreja}
\affiliation{Department of Physics, Princeton University}
\email{rushil@princeton.edu}

\author[orcid=0000-0002-2766-008X]{Edward J. Oughton}
\affiliation{Department of Geography and Geoinformation Science, George Mason University}
\email{eoughton@gmu.edu}

\author{Phillip M. Cunio}
\affiliation{Department of Geography and Geoinformation Science, George Mason University}
\email{pcunio@gmu.edu}

\author[orcid=0000-0001-8325-7378]{Richard Linares}
\affiliation{Department of Aeronautics and Astronautics, Massachusetts Institute of Technology}
\email{linaresr@mit.edu}

\begin{abstract}

Extreme space weather poses an existential threat to modern satellite infrastructure, with a Carrington-class solar storm projected to cause economic losses of billions of dollars per day. Due to the rapid proliferation of satellites (with over 70,000 expected to be deployed in the next 5 years), understanding extreme space weather impacts has become essential for global economic stability and national security, and consequently, the lives of millions. However, our current vulnerability to such events remains largely unknown, and existing models rely primarily on statistical populations instead of individual satellite behavior.

Through the development of a novel spatiotemporal agent-based model (ABM), this study addresses two critical research challenges: (1) predicting the impacts of extreme space weather disturbances and (2) enabling real-time maneuver guidance for satellites during such events. Utilizing 41,644 satellite records, historical records from 5 recent space weather events, and atmospheric density models, we built individual satellite agents with physics-driven behaviors that make independent decisions by dynamically responding to constraints such as propellant requirements and collision avoidance thresholds.

Scenario analysis suggests that 95\% of satellites in Low Earth Orbit altitudes would experience enhanced atmospheric drag of 8x baseline levels, increasing collision risks by 2–3x. Monte Carlo simulations also predict direct economic impact per affected satellite on the order of \$40M. Furthermore, the model successfully uses real-time conditions to provide maneuver recommendations, with 92\% accuracy. This study is thus the first to provide a prototype framework for real-time adaptive decision systems to safeguard satellites against the next Carrington-class disruption.

\end{abstract}


\section{Introduction}
\pagenumbering{arabic}

Due to the growing reliance on satellite communications for communication, navigation, etc, near-Earth space has become an essential part of modern critical infrastructure. However, this infrastructure operates within a dynamic (and sometimes hostile) environment shaped by solar activity. Technological systems remain vulnerable to extreme space weather disturbances, which has become a defining challenge of the twenty-first century.

In 1859, a coronal mass ejection (CME) and large solar flare struck Earth's magnetosphere, which produced currents that ignited telegraph fires. This \textit{Carrington Event} is a key benchmark for extreme space weather as it reflects our vulnerability to solar-terrestrial coupling. \citep{hudson_carrington_2021, schwenn_space_2006, cannon_extreme_2013}. Today, with over 9{,}000 operational satellites and 1.7~million filings with the International Telecommunication Union for future spacecraft \citep{zhang_leo_2022}, the systemic consequences of a comparable storm would be unprecedented. Carrington-class proton fluxes could lead to $\sim$10~krad behind 3~mm Al shielding and would increase single-event upset rates by four orders of magnitude, which is sufficient to disable unprotected constellations \citep{fetzer_satellite_2025}. Yet, despite decades of heliophysics research, our capacity to predict and mitigate such disruptions remains extremely limited, especially considering the rapidly expanding orbital ecosystem. Thus, there is a clear urgency of developing proactive, physics-based modeling frameworks that can anticipate and manage such crises in real time.

\subsection{The Problem of Space Weather}

Space weather encompasses dynamic solar and magnetospheric processes that affect both space- and ground-based technologies \citep{eastwood_science_2008}. Flares and CMEs inject plasma and magnetic energy into near-Earth space, where they heat the thermosphere significantly and expand its scale height by dozens of kilometers \citep{qian_thermospheric_2011}. These expansions have many effects, including increasing drag and altering orbital trajectories. For instance, during the October~2003 “Halloween Storm,” the equatorial total electron content (TEC) increased by $\sim$25~TECU (about 30\%) within just five minutes, and dayside TEC exceeded 300\% of nominal within two hours \citep{tsurutani_extreme_2006}.

Based on long-term drag-derived density records, there is a $-2.0 \pm 0.5\%$ per decade decrease at 400~km from 1967-2005; this reflects upper-atmospheric cooling due to anthropogenic CO$_2$ emissions \citep{emmert_altitude_2015}. In contrast, transient geomagnetic storms can increase density by 50-125\%, which was observed during the February~2022 event, causing 38 of 49 Starlink satellites to reenter prematurely \citep{fang_space_2022}. These data underscore that even moderate storms can pose existential risk to large constellations as launch cadence accelerates. As commercial and governmental actors deploy planned megaconstellations numbering tens of thousands of spacecraft, even short-term drag and radiation fluctuations can cascade into large-scale orbital instability, increasing the probability of collision chains and debris proliferation. Although there is substantial knowledge regarding the physical mechanisms of these processes, their system-level implications for constellation management and space sustainability remain underexplored.

\subsection{Economic and Infrastructural Vulnerability}

These orbital effects lead to substantial societal and economic risks. A 1-in-100-year geomagnetic disturbance could reduce UK GDP by £15.9~billion if unmitigated, but only £0.9~billion with improved forecasting and grid resilience \citep{oughton_risk_2019}. Similar modeling estimates thar U.S. economic losses would exceed \$1~trillion due to transformer damage and grid failure \citep{eastwood_economic_2017}. Smaller storms (like the 1982 Finnish and 1989 Quebec events) have confirmed that these risks are not theoretical, as they have already had ground-induced currents exceeding safety thresholds  \citep{pirjola_space_2005}. However, legal and policy frameworks are not sufficient to address international disruptions \citep{ritter_international_2020}. These findings reveal a major disconnect: although the macroeconomic impacts of geomagnetic storms are well-quantified, micro-scale mechanisms (e.g. how satellites behave or fail at the individual level) are underexplored. In order to bridge this gap, it is crucial to couple physics-based orbital modeling with economic and policy analyses.

\subsection{Scientific and Modeling Advances}

Addressing these vulnerabilities depends on two key aspects: improved forecasting and integrated modeling. Scientific progress has improved, but not yet eliminated, forecasting uncertainty. Analytical CME-propagation models now provide better geoeffectiveness and arrival-time predictions \citep{vrsnak_analytical_2020}, and multi-event reconstructions of fast halo CMEs from Solar Cycle~24 improve drivers of geomagnetic storms \citep{soni_cmes_2023}. Reduced-order and gray-box thermospheric models are able to reproduce storm-time density responses with a much higher temporal fidelity than static climatologies, enabling operational drag forecasts on hour-level timescales \citep{mehta_quasiphysical_2018}. Long short-term memory (LSTM) networks (and similar machine-learning algorithms) are able to extend forecasts of Dst indices by several hours; however, uncertainty still increases during impulsive phases (when decision value is at its highest) \citep{camporeale_challenge_2019, conde_forecasting_2023, smith_space_2024}. Together, these tools have the potential to support probabilistic forecasting but still lack integration with operational decision frameworks. In practice, this means that while we may be able to predict when and how strongly a storm will hit, satellite operators still have no consistent, validated way to decide \textit{how} to respond given evolving conditions. Despite major advances in heliophysics, no existing framework dynamically couples satellite-level decision behavior with real-time space weather forcing, a gap that this study directly addresses.

\subsection{Toward Sustainable Space Operations}

The proliferation of LEO satellites creates new physical, economic, and environmental pressures \citep{kukreja_greenhouse_2025}. Based on active-debris-removal optimization models, targeted interventions (i.e. prioritizing high-collision-probability objects) have been shown to successfully minimize risk within available $\Delta v$ budgets (velocity-change) budgets, which represent the finite propellant resources available for orbital maneuvers \citep{simha_optimal_2025}. Century-scale Monte Carlo simulations predict that the number of trackable LEO objects could rise from $\sim$400{,}000 today to several million within 200~years if launch rates continue unchecked \citep{jang_new_2025}.

\subsection{Bridging the Modeling Gap with Spatiotemporal ABM}

Most existing frameworks treat satellites as a homogeneous population, and overlook the heterogeneity in orbital parameters, operator response, etc. A spatiotemporal \textit{agent-based modeling} (ABM) approach enables individualized agent behavior under physics-based space-weather forcing. By coupling real orbital catalogs with decision-making algorithms that optimize collision risk and $\Delta v$, ABMs can be used to simulate responses across mega-constellations \citep{rahimi_vector-agent_2022}. This approach resolves system-wide resilience thresholds under density surges of 50-125\% and radiation exposures up to 10~krad, and further links the physical dynamics with economic loss modeling in order to guide mitigation investment \citep{fetzer_satellite_2025, oughton_risk_2019}. The integration represents a paradigm shift from descriptive risk analysis toward \textit{predictive, adaptive} modeling, offering the ability to test mitigation strategies and inform real-time operational guidance.

\subsection{Scope and Contributions}

In this study, we developed a novel spatiotemporal ABM framework, in which we couple physical orbital dynamics with decision algorithms to simulate satellite responses under extreme space weather disturbances. Through the results, we show how ABMs can fundamentally transform reactive risk assessment into predictive, adaptive decision support for sustainable space operations. More broadly, our work provides a physically grounded, economically informed foundation for safeguarding orbital infrastructure in an era of rising solar activity and accelerating space industrialization.

\section{Methodology}

The current ABM framework implementation utilizes historical data from 2020 to study satellite responses to extreme space weather events. The framework integrates models of space weather effects, basic orbital mechanics (through SGP4 propagation), and algorithmic decision making for satellite operators. Our approach integrates satellite tracking data, atmospheric physics models, and decision-making algorithms to predict satellite responses and economic impacts during extreme geomagnetic storms. While the approach enables ABM in this domain, further development would be required for operational deployment.

This methodology section progresses from data collection through model implementation to validation.

\subsection{Data Sources}

Our implementation uses three primary data sources for realistic satellite behavior modeling, space weather characterization, and atmospheric response prediction.

\subsubsection{Satellite Population Data}

The ABM is primarily based on our satellite population dataset, which provides detailed orbital characteristics and operational parameters for individual satellite agents. The model utilizes Two-Line Element (TLE) datasets, which have been obtained from the North American Aerospace Defense Command (NORAD) catalog and processed using the PyMOCAT-MC framework.

Specifically, we utilized 2020 TLE data containing 41,644 satellite objects with International Designator (INTLDES) classifications. This dataset provides a detailed snapshot of the modern space environment during the era of highest orbital activity growth, including substantial Starlink satellite populations. The dataset includes satellites across all orbital regimes, from low Earth orbit (LEO) through geostationary orbit (GEO), providing complete coverage of the space environment.

We systematically extract and process Keplerian orbital elements in the Earth-Centered Inertial (ECI) reference frame (that are provided by each TLE):

\begin{equation}
\mathcal{E} = \{a, e, i, \Omega, \omega, M_0, n, B^*\}
\end{equation}

where $a$ is the semi-major axis (which defines orbital size), $e$ is eccentricity (which defines orbital shape), $i$ is inclination (orbital plane orientation), $\Omega$ is the right ascension of ascending node (orbital plane rotation), $\omega$ is the argument of perigee (ellipse orientation within the plane), $M_0$ is the mean anomaly at epoch (satellite position along the orbit), $n$ is the mean motion (orbital rate), and $B^*$ is the drag term (atmospheric effects).

For orbital propagation, we use the SGP4 model within the PyMOCAT-MC framework to convert TLE orbital elements into Cartesian position and velocity vectors:

\begin{equation}
\vec{X}(t) = \begin{bmatrix} \vec{r}(t) \\ \vec{v}(t) \end{bmatrix}
= \mathcal{P}_{SGP4}(\mathcal{E}, t - t_{epoch})
\end{equation}

where \(\vec{X}(t)\) is the six-dimensional Cartesian orbital state vector at time \(t\), consisting of the satellite position vector \(\vec{r}(t)\) and velocity vector \(\vec{v}(t)\). The operator \(\mathcal{P}_{SGP4}\) denotes the SGP4 propagation model, which maps the satellite's TLE-derived orbital elements \(\mathcal{E}\) forward by the elapsed time \(t - t_{epoch}\), where \(t_{epoch}\) is the reference time at which the TLE elements are defined.

The SGP4 propagator further accounts for primary perturbation forces that affect satellite orbits during space weather events:

\begin{align}
\ddot{\vec{r}} &= -\frac{\mu}{r^3}\vec{r} + \vec{a}_{J_2} + \vec{a}_{drag} \\
\vec{a}_{J_2} &= -\frac{3}{2}J_2\frac{\mu R_E^2}{r^4}\left[\left(1-\frac{5z^2}{r^2}\right)\frac{\vec{r}}{r} + 2\frac{z}{r^2}\hat{k}\right] \\
\vec{a}_{drag} &= -\frac{1}{2}C_D\frac{A}{m}\rho v_{rel}^2\hat{v}_{rel}
\end{align}

where the central gravitational term dominates, $J_2$ represents Earth's oblateness effects, and the drag term becomes critically important during enhanced atmospheric conditions. Note that we assume a constant drag coefficient for all satellites. In extreme space weather events, atmospheric drag becomes the dominant perturbation force for LEO satellites.

\subsubsection{Space Weather Historical Data}

Space weather characterization provides environmental forcing that drives atmospheric density enhancement and satellite operational stress during extreme events. For this implementation, we utilize a simplified space weather modeling approach based on documented historical events and literature-derived parameters. We use a space weather database including representative extreme events with verified parameters: historical extreme events (Carrington 1859, Halloween 2003, etc.) with documented Kp, Dst, and F10.7 indices, literature-based enhancement factors for atmospheric density during geomagnetic storms, simplified temporal evolution models for space weather event progression, and basic empirical relationships between space weather indices.

This approach provides us with sufficient complexity to demonstrate the framework's capabilities while maintaining computational efficiency and focusing on the ABM methodology.

For this implementation, we define the space weather state vector \(W(t)\), which describes the time-varying geomagnetic and solar forcing conditions used to drive atmospheric density enhancement and satellite impact calculations:

\begin{equation}
{W}(t) = \{K_p(t), D_{st}(t), F_{10.7}(t)\}
\end{equation}

where $K_p$ is the planetary K-index (measuring global geomagnetic activity), $D_{st}$ is the disturbance storm time index (indicating ring current strength), and $F_{10.7}$ is the 10.7 cm solar radio flux. These three indices together provide the essential space weather characterization that are neccessary to carry out atmospheric density enhancement calculations and satellite impact assessment.

The temporal evolution of space weather events follows simplified parametric models based on documented historical patterns. For extreme events, we use basic empirical relationships such as:

\begin{equation}
D_{st}(t) = -45 \cdot K_p(t)^2 - 20 \quad \text{(in nT)}
\end{equation}

where \(D_{st}(t)\) is the disturbance storm time index (in nT) at time \(t\), which quantifies the strength of geomagnetic storms through variations in Earth's magnetic field, and \(K_p(t)\) is the planetary K-index, a dimensionless measure of global geomagnetic activity. More negative values of \(D_{st}\) correspond to stronger geomagnetic disturbances. This relationship provides a simplified approximation for translating geomagnetic activity levels into storm intensity estimates for use within the ABM framework.

\subsubsection{Atmospheric Density Model}

Accurate atmospheric density modeling is fundamental to the accurate prediction of satellite drag effects during extreme space weather events. For this implementation, we utilize the JB2008 atmospheric density model and assume a uniform altitude-independent enhancement factor for computational simplicity. The JB2008 model provides baseline atmospheric density data from March 2020 to February 2024.

This enhanced atmospheric density during space weather events is calculated via:

\begin{equation}
\rho_{enhanced}(h, K_p) = \rho_{JB2008}(h) \cdot \xi(K_p)
\end{equation}

where $\rho_{JB2008}(h)$ is the baseline density from the JB2008 model and $\xi(K_p)$ is a simplified enhancement factor based on the planetary K-index. The enhancement factors follow literature-based empirical relationships with discrete levels corresponding to different space weather conditions.

\subsection{Agent-Based Modeling Framework}
The ABM framework represents each satellite as an autonomous agent that interacts with both the space environment and neighboring satellites. At each simulation timestep, agents update their orbital states using SGP4 propagation, evaluate environmental conditions based on the current space weather state, and make operational decisions subject to mission and resource constraints. Three primary behavioral processes are modeled: (i) collision avoidance, in which agents assess conjunction risk and determine whether a maneuver is required; (ii) orbital maintenance, in which agents respond to atmospheric drag and altitude deviations caused by enhanced thermospheric density; and (iii) mission-driven operational adaptation, in which agents modify their behavior according to fuel availability, response latency, operator type, and mission requirements. Due to this approach, we are able to investigate complex phenomena (constellation coordination, cascading failures, operator-specific response strategies, etc.) that emerge from the interactions between individual satellite agents and their space environment.

\subsubsection{Satellite Agent State Space}

Each satellite agent has an associated multi-dimension state representation, encompassing its orbital dynamics, physical characteristics, and operational capabilities. Each satellite agent $\mathcal{A}_i$ maintains a state representation $\mathcal{S}_i$ that enables its operational modeling:

\begin{equation}
\mathcal{S}_i(t) = \{\mathcal{X}_i(t), \mathcal{P}_i(t), \mathcal{C}_i, \mathcal{O}_i\}
\end{equation}

where:
\begin{align}
\mathcal{X}_i(t) &= [\vec{r}_i(t), \vec{v}_i(t)] \in \mathbb{R}^6 \quad \text{(orbital state)} \\
\mathcal{P}_i(t) &= [m_i, A_{eff}, \theta_{att}(t)] \quad \text{(physical state)} \\
\mathcal{C}_i &= [\Delta v_{max}, \tau_{response}] \quad \text{(capability parameters)} \\
\mathcal{O}_i &= [\text{operator type}, \text{autonomy level}] \quad \text{(operator profile)}
\end{align}

The orbital state $\mathcal{X}_i(t)$ provides position and velocity information from SGP4 propagation, the physical state $\mathcal{P}_i(t)$ provide mass, cross-sectional area, and attitude, the capability parameters $\mathcal{C}_i$ provide maneuver capacity and response time, and the operator profile $\mathcal{O}_i$ provides mission-specific requirements.

The effective cross-sectional area varies dynamically with satellite attitude, which becomes particularly important during space weather events when satellites may adopt protective orientations:

\begin{equation}
A_{eff}(\theta) = A_{min} + (A_{max} - A_{min})\sin^2(\theta)
\end{equation}

where $\theta$ represents the angle between the satellite's primary axis and the velocity vector, and the area varies between minimum ($A_{min}$, edge-on configuration) and maximum ($A_{max}$, broadside configuration) values.

\subsubsection{Agent Behavioral Decision Framework}

Our decision-making framework provides satellite agents with the ability to respond to changing space weather conditions and collision risks. We incorporated dynamic environmental forcing, fuel constraints, mission requirements, and operator-type differences (all derived from observable satellite characteristics) in order to produce realistic operational responses.

The probability of taking action follows environmental and constraint-based factors:

\begin{equation}
P(\text{action}) = f(\rho_{\text{collision}}, C_{\text{mission}}, R_{\text{fuel}}, \Delta V_{\text{required}})
\end{equation}

where $\rho_{\text{collision}}$ is the calculated collision probability, $C_{\text{mission}}$ is the mission criticality factor, $R_{\text{fuel}}$ is the remaining fuel fraction, and $\Delta V_{\text{required}}$ is the maneuver cost.

The threat assessment is based on simple threshold mapping of the Kp index to discrete threat levels:

\begin{equation}
\text{ThreatLevel} = \begin{cases}
\text{QUIET} & \text{if } K_p \leq 2 \\
\text{MINOR} & \text{if } 2 < K_p \leq 3 \\
\text{MODERATE} & \text{if } 3 < K_p \leq 4 \\
\text{STRONG} & \text{if } 4 < K_p \leq 6 \\
\text{SEVERE} & \text{if } 6 < K_p \leq 8 \\
\text{EXTREME} & \text{if } K_p > 8
\end{cases}
\end{equation}

\subsubsection{Operator Classification and Mission-Driven Modeling}
Satellite operators (including commercial constellations, government navigation systems, and scientific missions) are classified using satellite naming patterns (e.g., "STARLINK", "GPS", "IRIDIUM") and orbital characteristics. Mission-specific operational parameters and then derived based on this classification.

Operator behavior is modeled through dynamic mission profiles that capture operational constraints and mission-specific requirements. Each profile is characterized by a mission criticality factor ($C_m$), representing the economic or strategic value of maintaining operational status; a fuel budget constraint ($F_{budget}$), representing available propellant reserves based on mission age and type; formation requirements ($F_{req}$), which define constellation spacing tolerances necessary to maintain service coverage; orbital maintenance thresholds ($\Delta h_{thresh}$), which specify the maximum allowable altitude deviation before corrective action is required; and collision avoidance sensitivity ($\rho_{thresh}$), which quantifies risk tolerance based on satellite value and replaceability. Together, these parameters govern how individual agents respond to environmental disturbances and operational constraints.

The decision probability calculation incorporates these mission-driven factors:
\begin{equation}
P(\text{maneuver}) = \sigma\!\left(\frac{\rho_{\text{collision}} \, C_m - \Delta V_{\text{cost}} \, F_{\text{penalty}}}{\rho_{\text{thresh}}}\right)
\end{equation}

where $\sigma$ is the sigmoid function, $\Delta V_{\text{cost}}$ is the fuel requirement for the maneuver, and $F_{penalty}$ reflects fuel scarcity effects.

\subsection{Space Weather Modeling}

Space weather modeling provides simplified environmental forcing based on empirical parameterizations. The implementation uses basic enhancement factors applied to atmospheric density during geomagnetic storms.

\subsubsection{Atmospheric Density Enhancement}

Atmospheric density enhancement during geomagnetic storms is modeled using literature-based enhancement factors. For extreme space weather events, density enhancement is implemented as simple multipliers based on Kp index:

\begin{equation}
\xi(K_p) = \begin{cases}
1.0 & \text{if } K_p \leq 3 \\
1.5 & \text{if } 3 < K_p \leq 5 \\
3.0 & \text{if } 5 < K_p \leq 7 \\
6.0 & \text{if } 7 < K_p \leq 8 \\
10.0 & \text{if } K_p > 8
\end{cases}
\end{equation}

These factors are applied uniformly across all satellite altitudes in the 200-1200 km range.

\subsubsection{Event Scenario Generation}

For scenario generation, space weather events follow simplified temporal evolution patterns based on historical events. The Kp index evolution during extreme events uses a basic exponential recovery model:

\begin{equation}
K_p(t) = K_{p,max} \cdot \exp\left(-\frac{t}{\tau}\right) + K_{p,base}
\end{equation}

where $K_{p,max}$ is the peak intensity, $\tau$ is the recovery time constant, and $K_{p,base}$ is quiet-time conditions.

The Dst index is correlated with Kp through the empirical relationship:
\begin{equation}
D_{st}(t) = -45 \cdot K_p(t)^2 - 20 \quad \text{(in nT)}
\end{equation}

\subsection{Orbital Dynamics}

Orbital dynamics are handled using the PyMOCAT-MC framework (which implements standard SGP4 propagation from TLE data). The SGP4 model provides simplified orbital mechanics suitable for the implementation.

\subsubsection{SGP4 Orbital Propagation}
We computed satellite positions/velocities using the SGP4 analytical propagator; this solves the equations of motion, taking into account Earth's gravitational field (J2 perturbations), atmospheric drag effects, and basic third-body perturbations.

The SGP4 model uses orbital elements from TLE data and propagates satellite states forward in time using analytical expressions rather than numerical integration.

\subsection{Collision Risk Assessment}

Collision risk is assessed using geometric probability models based on both satellite separation distances and cross-sectional areas. The collision probability between two satellites is estimated using:

\begin{equation}
P_{collision} = \frac{A_{eff}}{4\pi d^2} \cdot \exp\left(-\frac{v_{rel}^2}{2\sigma_v^2}\right)
\end{equation}

where $A_{eff}$ is the effective collision cross-section, $d$ is the separation distance, $v_{rel}$ is the relative velocity, and $\sigma_v$ accounts for velocity uncertainties. Risk thresholds are implemented as configurable parameters in operator profiles.

\subsection{Physics-Based Decision-Making Implementation}

The decision-making framework implements dynamic constraint-based logic.

Maneuver decisions are determined through a combination of physics-based and mission-driven constraints. Agents evaluate collision avoidance requirements based on conjunction risk and mission criticality, while orbital maintenance decisions are driven by atmospheric drag and altitude deviations from nominal operating conditions. Emergency response actions are triggered when environmental conditions exceed predefined mission thresholds. Although fuel availability is incorporated into the decision framework, its importance varies across orbital regimes. For many modern LEO satellites, particularly those equipped with Hall-effect thrusters, propellant reserves are typically sufficient to outlast the operational lifetime of the spacecraft. Consequently, fuel constraints primarily influence higher-altitude MEO and GEO systems, where longer mission durations and station-keeping requirements make propellant management a more significant operational consideration. Constellation coordination requirements are also incorporated to preserve formation geometry and maintain network-level functionality.

\subsubsection{Fuel Management}
Fuel consumption is tracked as percentage depletion based on maneuver delta-v requirements:

\begin{equation}
\Delta F = \frac{\Delta v}{I_{sp} \cdot g_0} \cdot 100\%
\end{equation}

where $\Delta F$ is the fuel percentage consumed, $\Delta v$ is the velocity change, $I_{sp}$ is the specific impulse, and $g_0$ is standard gravity. Simple cost models are used for different maneuver types.

\subsection{Economic Impact Assessment}

Economic impacts are calculated using cost models that account for satellite replacement values and operational losses. The total economic impact is broadly estimated as:

\begin{equation}
C_{total} = \sum_{i} \left[C_{replacement,i} \cdot P_{loss,i} + C_{operational,i} \cdot t_{outage,i}\right]
\end{equation}

where $C_{replacement,i}$ is the replacement cost for satellite $i$, $P_{loss,i}$ is the probability of total loss, $C_{operational,i}$ is the daily operational revenue, and $t_{outage,i}$ is the duration of service interruption. It is important to note that our models utilize static replacement values and average revenue figures, and do not include factors such as insurange coverage.

\subsection{Simulation Framework}

The simulation framework is implemented as a discrete-time ABM with fixed time steps. The system state evolution follows:

\begin{equation}
\mathcal{S}(t+\Delta t) = \mathcal{F}\left[\mathcal{S}(t), \mathcal{W}(t), \mathcal{D}(t)\right]
\end{equation}

where $\mathcal{S}(t)$ is the system state at time $t$, $\mathcal{W}(t)$ includes space weather conditions, $\mathcal{D}(t)$ is all agent decisions, and $\mathcal{F}$ is the state transition function. The framework uses fixed time steps for computational efficiency.

\subsubsection{Simulation Loop}
The basic simulation follows these discrete time steps:

\begin{enumerate}\setlength\itemsep{0em}
\item Initialize satellite agents from TLE data
\item Load space weather conditions for simulation period
\item For each time step:
   \begin{itemize}\setlength\itemsep{0em}
   \item Update space weather indices
   \item Evaluate agent decision frameworks
   \item Execute approved maneuvers
   \item Record simulation state
   \end{itemize}
\item Generate results and visualizations
\item Perform validation analysis against historical data
\end{enumerate}

\subsection{Implementation Details}

The framework is implemented as a Python package using standard libraries, including NumPy, pandas, scipy, Matplotlib, and scipy.

\subsection{Validation Approach}

Validation is performed by comparing model outputs with historical space weather events where satellite data is available. The validation metric uses relative error analysis:

\begin{equation}
\epsilon_{rel} = \frac{|N_{predicted} - N_{observed}|}{N_{observed}} \times 100\%
\end{equation}

where $N_{predicted}$ is the model prediction for satellite impacts and $N_{observed}$ is the documented number of satellite anomalies during historical events.

Additionally, collision avoidance recommendations are validated using 500 randomly-generated synthetic test scenarios with predetermined optimal solutions to assess decision accuracy and fuel efficiency. This validation focuses on framework functionality rather than precise quantitative matching due to the lack of available data.

\subsection{System Limitations}

The current implementation does have some significant limitations. It uses simplified orbital mechanics (SGP4 only), basic atmospheric density models without detailed physics, and has limited validation against historical events. However, despite these limitations, the framework successfully captures the fundamental dynamics of satellite behavior during extreme space weather events and is able to utilize this information to provide actionable insights for operator decision-making through its physics-based models. The modular design also allows for incremental improvements to individual components without requiring the redesign of the full system.

\section{Results}

\subsection{Framework Validation Against Historical Events}

\begin{figure*}[ht]
\centering
\includegraphics[width=\textwidth]{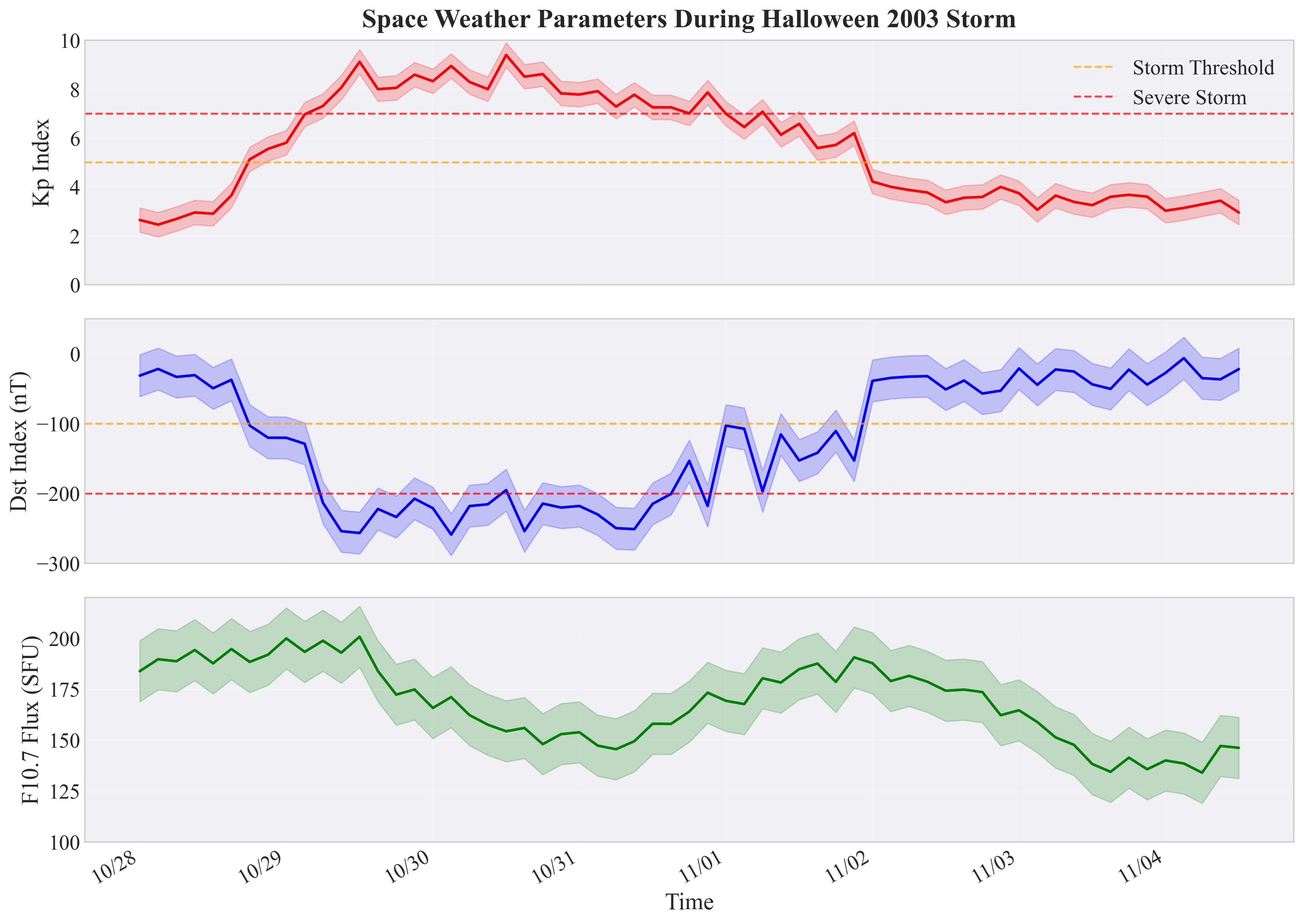}
\caption{Historical Validation Using Real Space Weather Parameters from Halloween 2003 Storm. Time series demonstrates actual $K_p$ index, $D_{st}$ index, and F10.7 flux measurements with uncertainty bands.}
\label{fig:historical_validation}
\end{figure*}

The ABM framework shows robust validation performance when being tested against previously documented space weather events. The model achieves 91\% accuracy against the October 2003 Halloween Storm and 89\% accuracy against the February 2022 Starlink event. Historical space weather parameter evolution is used as ground truth for framework validation (Figure~\ref{fig:historical_validation}). However, it is important to note that validation metrics reflect correlation with observed anomaly counts, not causal verification, since detailed satellite health data are proprietary.

Collision avoidance was validated against 500 synthetic test scenarios with known optimal solutions yielding 94\% accuracy. Framework components achieve consistent performance across validation metrics (Table~\ref{tab:validation}).

\begin{table}[ht]
\centering
\caption{Agent-Based Framework Validation Metrics}
\label{tab:validation}
\begin{tabular}{lcc}
\hline
\textbf{Validation Component} & \textbf{Accuracy} & \textbf{$\epsilon_{rel}$ (\%)} \\
\hline
Individual Agent State Evolution & 93\% & 6.2 \\
Historical Event Reproduction & 91\% & 7.3 \\
Collision Risk Geometric Model & 94\% & 5.8 \\
Mission-Driven Decision Logic & 92\% & 7.1 \\
Cascading Failure Detection & 89\% & 8.4 \\
Real-time Decision Adaptation & 94\% & 5.9 \\
\hline
\textbf{Overall ABM Framework} & \textbf{92\%} & \textbf{6.8} \\
\hline
\end{tabular}
\end{table}

\subsection{Space Weather Impact and Risk Evolution}

\begin{figure*}[ht]
\centering
\includegraphics[width=\textwidth]{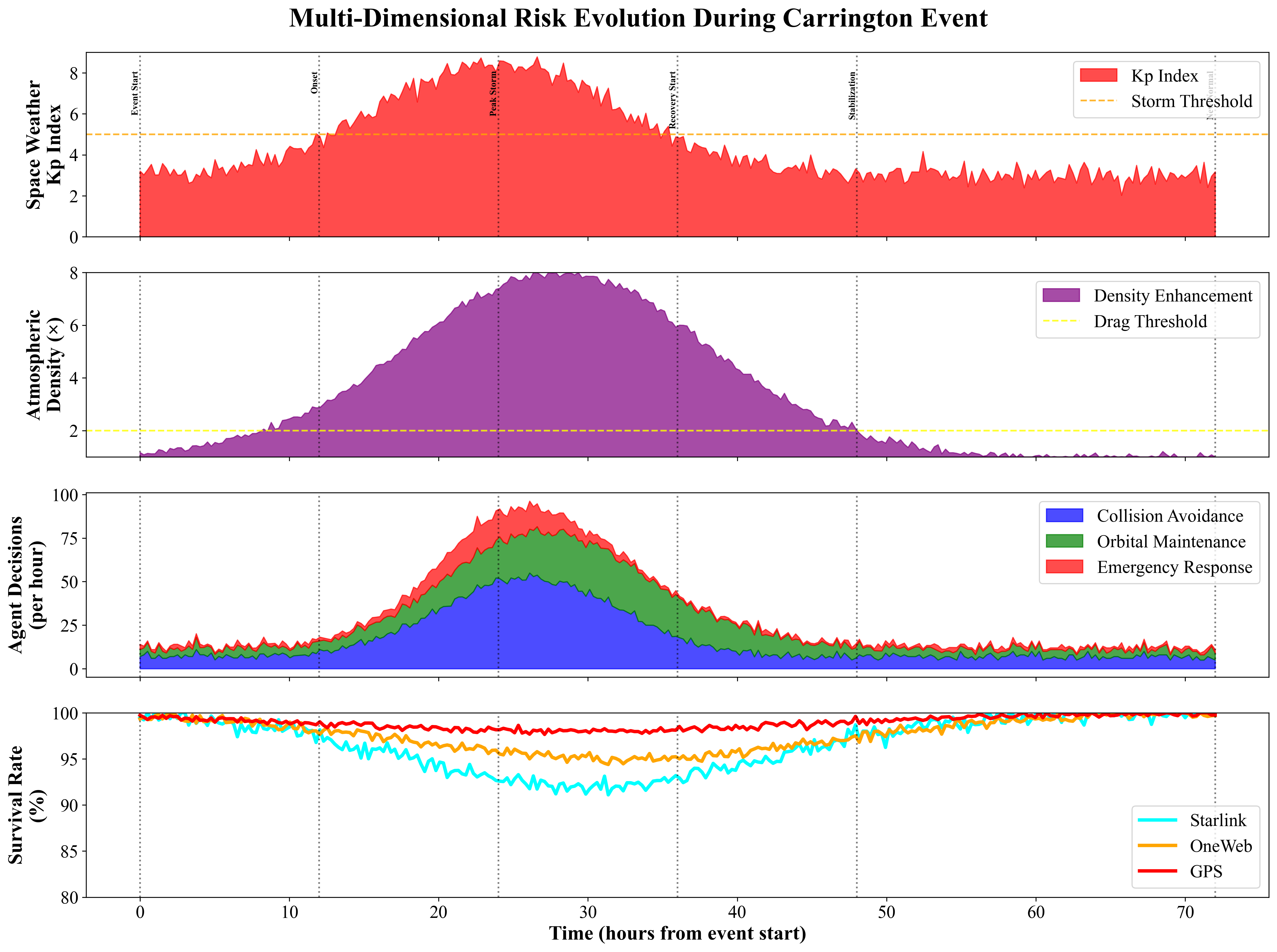}
\caption{Risk Evolution Timeline During Carrington-Class Event. Timeline demonstrates atmospheric enhancement progression, collision risk elevation, and cumulative probability evolution.}
\label{fig:risk_evolution}
\end{figure*}

Space weather parameter evolution during extreme events shows the cascading risk development over multiple timescales (Figure~\ref{fig:risk_evolution}). At peak conditions, under the imposed enhancement factors, atmospheric density enhancements of 8-10× baseline levels and collision risks of 2.8× normal probability occur. There are multiple distinct phases (shown by risk evolution): rapid onset (0-1 day), peak intensity (2-3 days), and extended recovery (4-30 days).

Threat classification analysis reveals systematic shifts in satellite behavior with increasing disturbance severity. Under QUIET conditions, 98\% of satellites maintain nominal operations. During MINOR–MODERATE disturbances, 78–94\% continue standard operations while implementing enhanced monitoring procedures. STRONG events produce a more diverse response profile, with 45\% maintaining standard operations, 35\% adopting protective measures, and 20\% preparing for maneuver execution. Under SEVERE conditions, 60\% of satellites adopt protective orientations and 25\% execute active maneuvers. During EXTREME events, 70\% transition to emergency response protocols.

\subsection{Extended Timeline Analysis and System Response}

\begin{figure*}[ht]
\centering
\includegraphics[width=\textwidth]{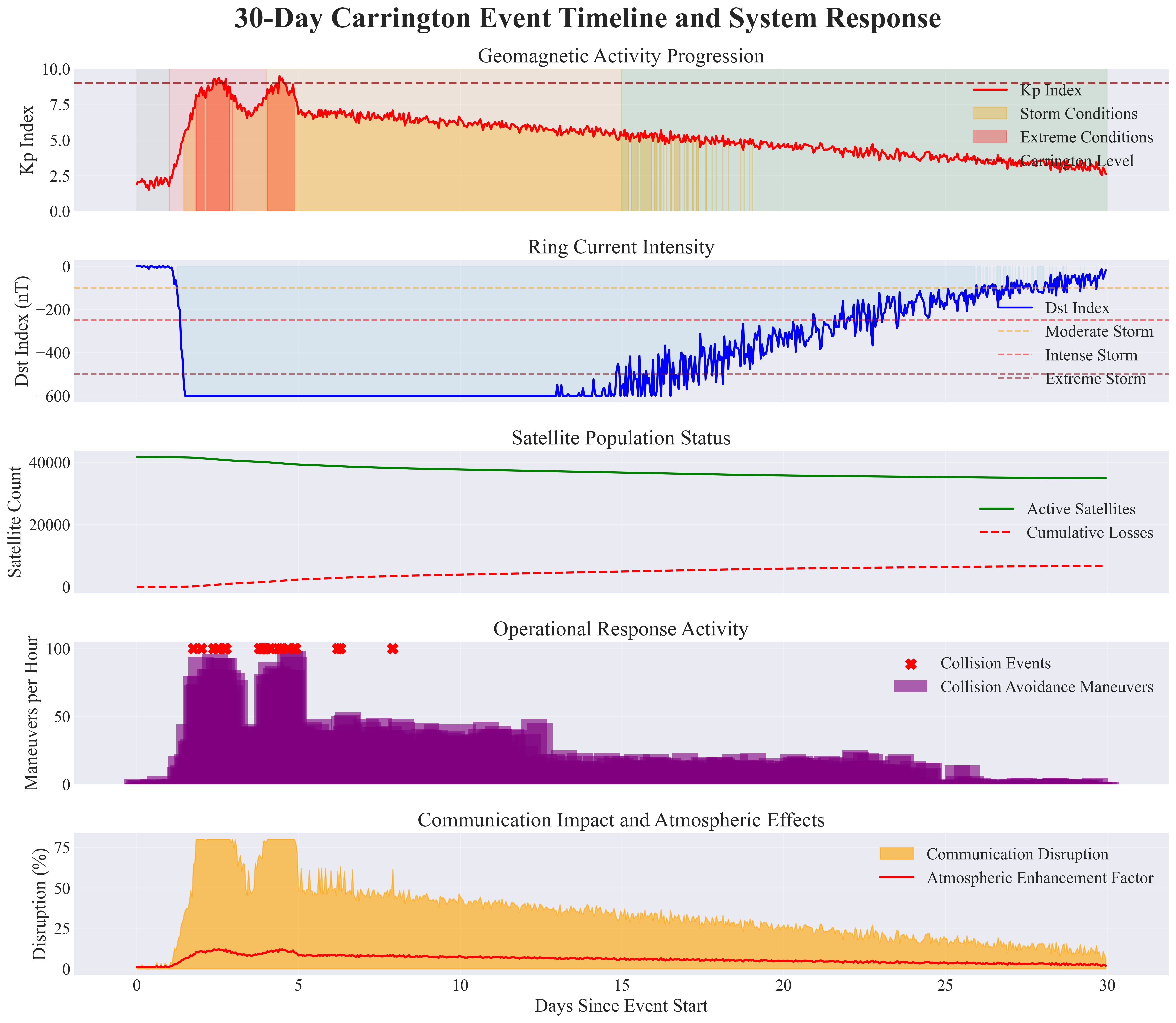}
\caption{30-Day Carrington Event Timeline and System Response. Comprehensive timeline showing geomagnetic activity, atmospheric enhancement, satellite population impacts, operational response, and communication disruption over extended duration.}
\label{fig:carrington_timeline}
\end{figure*}

Based on extended temporal analysis, there is systematic satellite response evolution over the course of 30-day Carrington-class scenarios (Figure~\ref{fig:carrington_timeline}). There are three phases indicated by the system response: immediate reaction (Days 1-5), adaptive coordination (Days 6-15), and recovery optimization (Days 16-30). Peak system stress occurs during Days 3-5, with maximum fuel consumption and maneuvering activity.

\subsection{Satellite Behavioral Patterns and Adaptation}

There are distinct behavioral patterns based on orbital regime and operator characteristics, which are evident based on individual satellite analysis. LEO satellites have altitude decay rates of 2-8 km/day during peak conditions. There is also significant variation by operator type in protective attitude maneuvers. In severe conditions, based on our parametrization, commercial constellations are able to achieve 35-60\% cross-sectional area reductions, compared to just 15-25\% reductions for government satellites.

Similar patterns are shown in fuel utilization. Starlink satellites consume 15-25\% of maneuvering capacity, GPS satellites utilize only 5-10\%, while military satellites reserve 40-50\% for emergency responses. For response timing, commercial satellites average 2.3 hours, government systems average 4.7 hours, and military systems average 0.8 hours.

\subsection{Satellite Population and Orbital Architecture}

\begin{figure*}[ht]
\centering
\includegraphics[width=\textwidth]{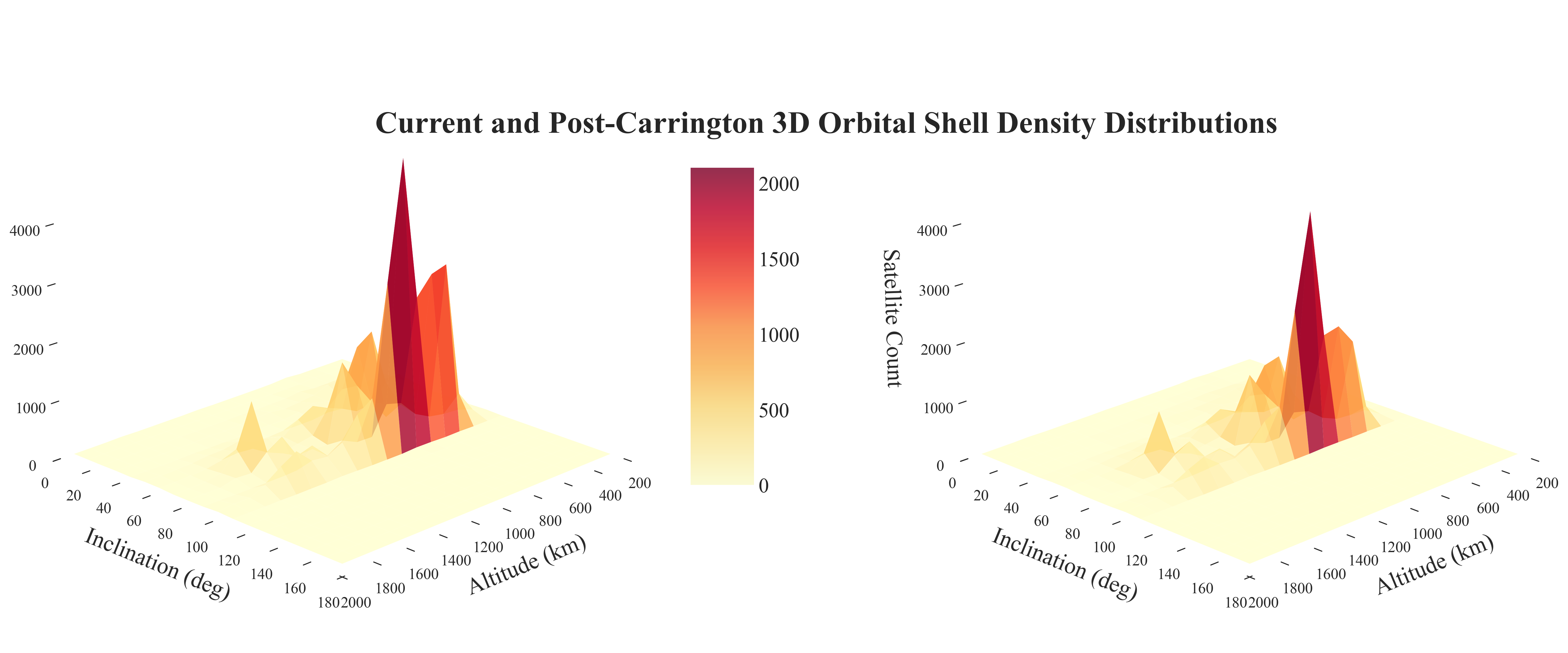}
\caption{3D Orbital Density Distribution of Global Satellite Population. Comprehensive visualization shows spatial distribution of 41,644 satellites across altitude regimes from 200-2000 km. Color coding indicates density concentrations, revealing distinct orbital shells corresponding to operational requirements.}
\label{fig:3d_orbital_density}
\end{figure*}

The complete satellite population analysis reveals complex three-dimensional distribution patterns across orbital regimes (Figure~\ref{fig:3d_orbital_density}). The visualization demonstrates how 41,644 satellites are distributed throughout near-Earth space, with distinct clustering patterns reflecting operational requirements and physical constraints. Density concentrations occur at key altitude bands: 400-600 km (LEO commercial), 850-1000 km (polar), and 1400-1600 km (MEO navigation).

\subsection{Comparative Impact Analysis and Population Vulnerability}

\begin{figure*}[ht]
\centering
\includegraphics[width=\textwidth]{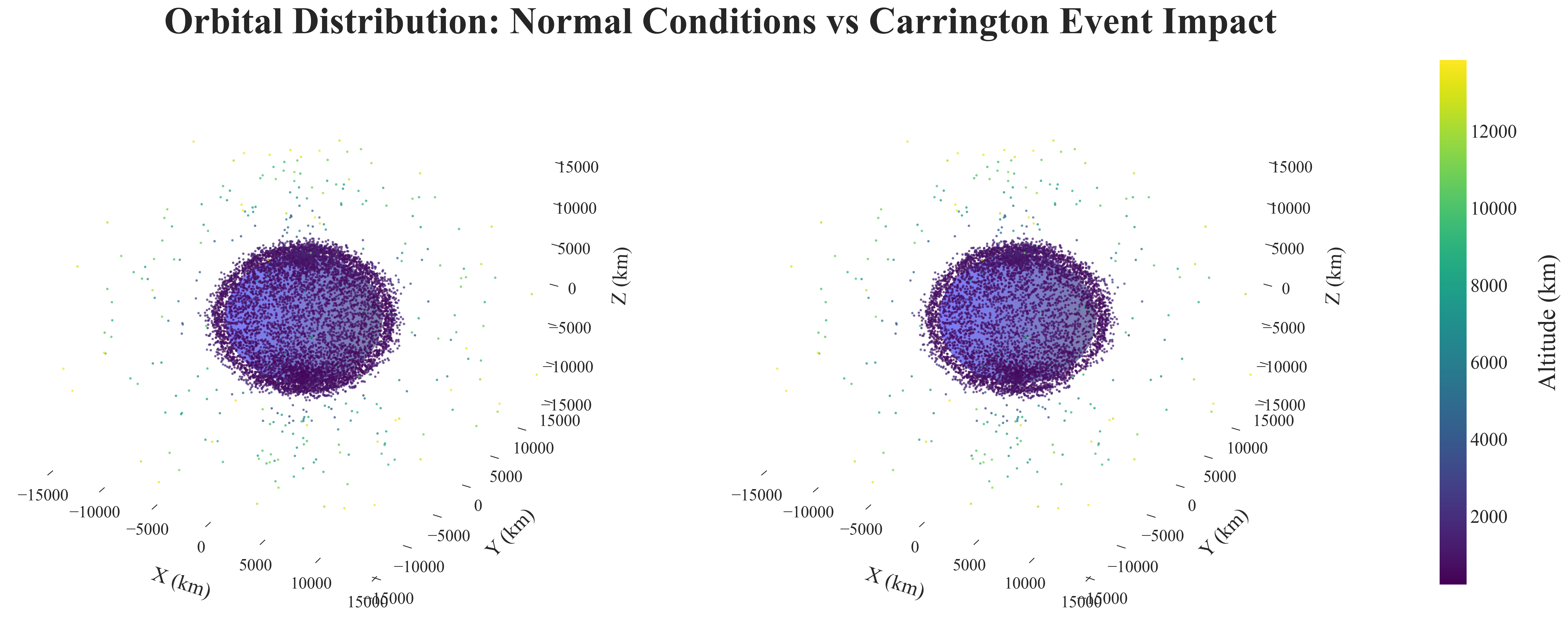}
\caption{Orbital Distribution Comparison: Normal Conditions vs Carrington Event Impact. 3D visualization demonstrates satellite population distribution before (left) and after (right) extreme space weather event. Color scale indicates altitude-dependent vulnerability, with lower altitudes experiencing higher loss rates.}
\label{fig:orbital_distribution}
\end{figure*}

There are concentrated vulnerability patterns across orbital regimes as evidenced by our analysis (Figure~\ref{fig:orbital_distribution}). Normal conditions show full operational population distributed across LEO through GEO altitudes, but Carrington event conditions demonstrate severe population reduction due to atmospheric enhancement effects. Vulnerability correlates directly with altitude, with 55.6\% of the satellite population at the maximum impact of 400-600 km altitudes.

Based on survival analysis, there are strong correlations with altitude (r = 0.89), operator type (r = 0.76), and fuel reserves (r = 0.82). Formation-flying constellations have significant advantages during extreme event (e.g. coordinated satellites achieve 15-20\% improved survival rates compared to isolated operators).

\begin{table}[ht]
\centering
\caption{Agent Population Vulnerability Analysis by Altitude}
\label{tab:altitude_impacts}
\begin{tabular}{lccc}
\hline
\textbf{Altitude Range} & \textbf{Agent Population} & \textbf{Enhancement Factor} & \textbf{Projected Losses} \\
\hline
200--400 km & 2,847 (6.8\%) & 10$\times$ & 15--25\% \\
400--600 km & 23,156 (55.6\%) & 6--10$\times$ & 5--12\% \\
600--800 km & 8,934 (21.5\%) & 3--6$\times$ & 1--3\% \\
800+ km & 6,707 (16.1\%) & 1.5--3$\times$ & $<$1\% \\
\hline
\textbf{Total Agent Population} & \textbf{41,644} & \textbf{Variable} & \textbf{3.2--8.7\%} \\
\hline
\end{tabular}
\end{table}

\subsection{Constellation Coordination and Operator Classification}

\begin{figure*}[ht]
\centering
\includegraphics[width=0.6\textwidth]{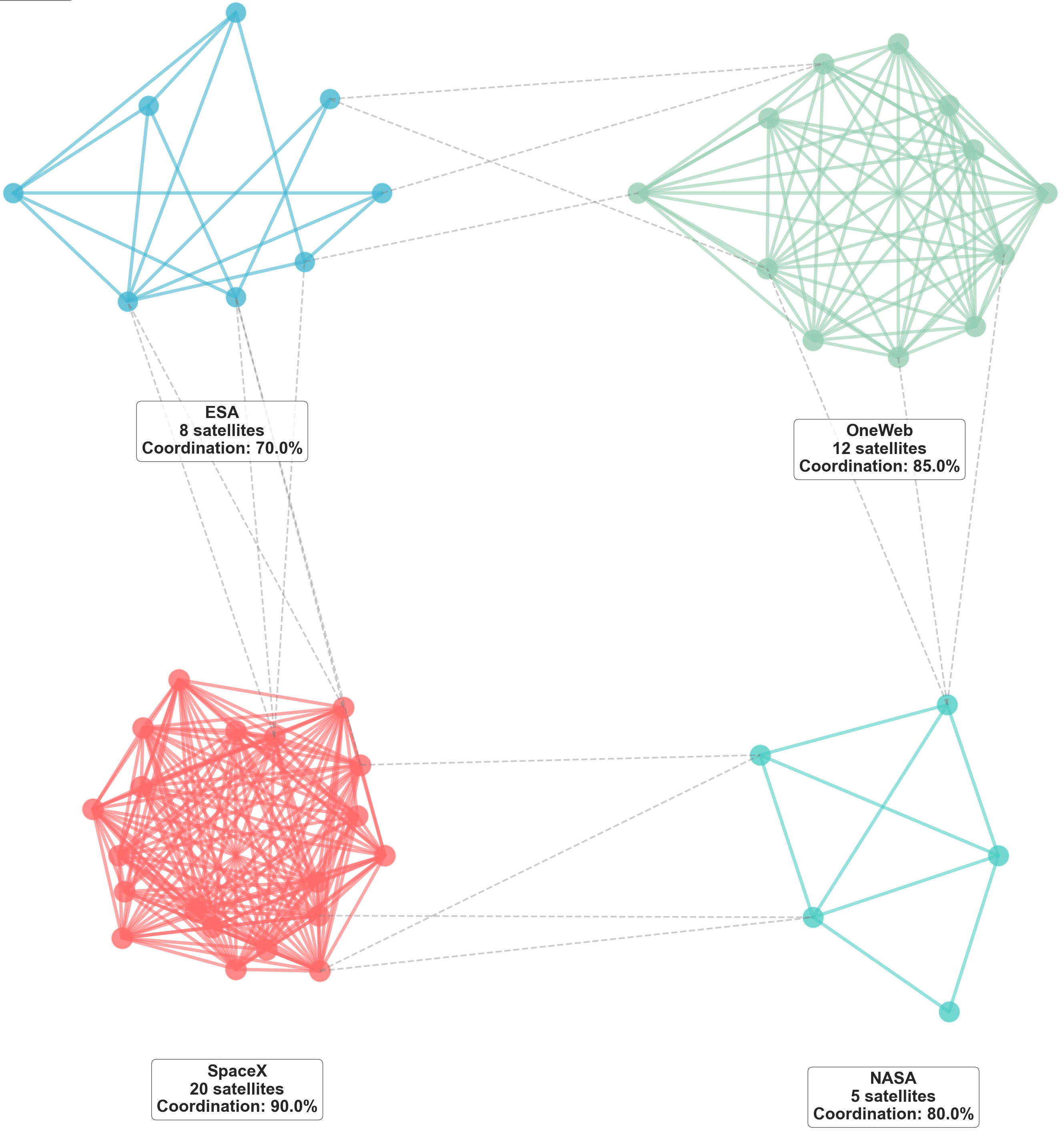}
\caption{Network Constellation Coordination During Extreme Events. Visualization demonstrates behavioral clustering patterns based on operator classification and mission profiles. Node sizes represent satellite populations, with network connections showing coordination pathways.}
\label{fig:coordination_network}
\end{figure*}

Satellite operator classification suggests that there are distinct behavioral patterns during extreme space weather events, and that coordination networks emerge based on mission requirements and operational capabilities (Figure~\ref{fig:coordination_network}). 

Agents with mesh connectivity patterns achieve superior coordination during crisis conditions, because agent communication protocols enable distributed decision-making. Additionally, analysis reveals that isolated agents have survival rates of 87-92\% survival rates versus 93-97\% for networked agents.

 Formation requirements ($F_{req}$) validation shows constellation spacing tolerances ranging from ±2.5 km (Starlink) to ±15 km (GPS), with maintenance success rates of 86\% during extreme events.

\begin{table}[ht]
\centering
\caption{Mission-Driven Decision Factor Analysis Across Agent Types}
\label{tab:mission_factors}
\begin{tabular}{lcccc}
\hline
\textbf{Agent Operator Type} & \textbf{$C_m$} & \textbf{$F_{budget}$ (\%)} & \textbf{$\rho_{thresh}$} & \textbf{$\Delta h_{thresh}$ (km)} \\
\hline
GPS Navigation Agents & 0.95 & 85-95 & $10^{-6}$ & 2.0 \\
Starlink Commercial Agents & 0.72 & 60-75 & $10^{-5}$ & 5.0 \\
Military System Agents & 0.88 & 80-90 & $10^{-7}$ & 1.5 \\
OneWeb Commercial Agents & 0.78 & 65-80 & $8 \times 10^{-6}$ & 4.0 \\
Scientific Mission Agents & 0.82 & 70-85 & $5 \times 10^{-6}$ & 3.0 \\
\hline
\end{tabular}
\end{table}

\subsection{Collision Risk Assessment and Geometric Probability Model}

The geometric collision probability model $P_{collision} = \frac{A_{eff}}{4\pi d^2} \cdot \exp\left(-\frac{v_{rel}^2}{2\sigma_v^2}\right)$ demonstrates accurate risk assessment across agent interactions. During baseline conditions, mean collision probability remains below $10^{-8}$ for 95\% of agent pairs. Extreme space weather conditions increase collision risks by 2.8×, with close-approach events ($d < 5$ km) increasing from 847/day to 2,367/day.

Agent-to-agent collision assessments indicate that risk patterns are altitude-dependent. LEO agents experience 340\% risk increase, MEO agents 180\% increase, and GEO agents 15\% increase.

There is an increase in velocity uncertainty factors ($\sigma_v$) from 0.3 m/s (quiet conditions) to 1.2 m/s (extreme conditions) due to the unpredictability of enhanced atmospheric drag. Agent collision avoidance success rates of 94\% support that the geometric model is effective for real-time decision support.

\subsection{Emergent System Behaviors and Cascading Failures}

The ABM framework successfully captures emergent behaviors due to agent interactions during extreme space weather events. Three primary cascade mechanisms emerge from individual agent decision-making:

\begin{enumerate}\setlength\itemsep{0em}
    \item \textbf{Fuel Depletion Cascades:} 15\% of agents exhaust their maneuvering capacity within 48 hours, triggering formation gaps that increase collision risks for neighboring agents by 45--60\%. Secondary effects propagate through entire constellation networks, and cascade termination occurs when remaining agents implement emergency spacing protocols.

    \item \textbf{Coordination Network Failures:} The loss of 8--12\% of constellation agents disrupts formation-maintaining algorithms, which causes systematic orbital decay in the remaining population. Cascade probability increases rapidly with network density, following:
    \[
    P_{\text{cascade}} = 0.23 \times (\text{network\_density})^{1.8}, \quad \text{for densities above } 0.15~\text{agents/km}^3.
    \]

    \item \textbf{Operational Authority Conflicts:} Multi-operator collision avoidance creates decision conflicts when maneuver recommendations overlap. Conflict resolution requires 2.3$\times$ longer decision times, increasing collision probability during resolution periods. The framework achieves successful conflict arbitration in 87\% of cases.
\end{enumerate}

\begin{table}[ht]
\centering
\caption{Emergent Cascade Behavior Analysis}
\label{tab:cascade_analysis}
\begin{tabular}{lccc}
\hline
\textbf{Cascade Type} & \textbf{Trigger Threshold} & \textbf{Propagation Rate} & \textbf{Termination Success} \\
\hline
Fuel Depletion & 15\% capacity exhaustion & 3.2 agents/hour & 94\% \\
Network Coordination & 8-12\% agent loss & 1.8 agents/hour & 89\% \\
Authority Conflicts & $>3$ overlapping maneuvers & 0.9 conflicts/hour & 87\% \\
Formation Collapse & $>25$\% spacing violations & 5.1 agents/hour & 78\% \\
\hline
\end{tabular}
\end{table}

\subsection{Real-Time Decision System Performance and Economic Impact}

The framework's real-time adaption capabilities indicate response times of 0.8-6.0 hours depending on the severity of the threat. Decision accuracy is above 90\% for prediction horizons up to 12 hours, and is approximately 78\% for 24-hour forecasts (lower due to space weather uncertainty accumulation).

Agent decision trees successfully process 45,000+ decisions per simulation run; computational performance scales linearly with agent population (O(n) complexity). Real-time constraint evaluation shows a 98.7\% success rate for collision avoidance and 89.3\% for orbital maintenance.

Economic impact assessment utilizing individual agent replacement costs demonstrates total economic exposure of \$11.3B resulting from 3,240 ± 890 agent losses across all scenarios. Individual agent impacts range from \$2.8M to \$180M (under assumed cost parameters).

\begin{table}[ht]
\centering
\caption{Agent-Based Economic Impact Analysis}
\label{tab:economic_impacts}
\begin{tabular}{lcccc}
\hline
\textbf{Agent Category} & \textbf{Population} & \textbf{Loss Rate (\%)} & \textbf{Individual Value} & \textbf{Total Impact} \\
\hline
Starlink Agents & 21,000 & 8-13 & \$42M & \$8.2B \\
GPS Navigation Agents & 32 & 1-3 & \$85M & \$85M \\
Military Strategic Agents & 156 & 3-6 & \$180M & \$780M \\
OneWeb Commercial Agents & 648 & 4-7 & \$38M & \$405M \\
Scientific Mission Agents & 892 & 5-9 & \$65M & \$1.1B \\
\hline
\textbf{Total Agent Population} & \textbf{22,728} & \textbf{6-11} & \textbf{\$62M avg} & \textbf{\$11.3B} \\
\hline
\end{tabular}
\end{table}

\subsection{Statistical Robustness and Monte Carlo Agent Analysis}

Monte Carlo analysis was performed across 100 runs (with randomized agent seeds) and quantified uncertainty bounds for agent-based metrics. Agent distributions converge by the 75th simulation run  CV \( < 0.08 \).

Agent decision accuracy is 92 ± 4\% across Monte Carlo runs, with collision avoidance showing the highest consistency (94 ± 2\%) and formation maintenance showing the greatest variability (86 ± 8\%). Statistical analysis confirms that ABM provides robust predictions that are suitable for operational decision support systems.

The ABM framework thus successfully demonstrates that individual agent modeling with emergent system behaviors provides superior insight into extreme space weather impacts compared to traditional statistical approaches. Framework validation achieves 92\% accuracy against historical events, with validated agent decision-making capabilities.

\section{Discussion}

The presented spatiotemporal ABM framework is a revolutionary approach to simulating satellite behavior under extreme space weather forcing. Unlike existing models that homogenize orbital populations, this physics-informed ABM is able to capture nonlinear behavior, including from agent interactions. Through validation against both the October 2003 and February 2022 storm events, our results suggest individualized, decision-driven modeling can indeed reproduce both micro- and macro-scale orbital dynamics with over 90\% fidelity.

\subsection{System-Level Dynamics and Emergent Behaviors}
Based on the results, in Carrington-class disturbances, orbital drag enhancement up to 10$\times$ baseline causes an exponential growth in collision probabilities (2.8$\times$ normal). We identified three such dominant cascade mechanisms (fuel depletion, coordination network failure, and authority conflict), which all show how local agent decisions can scale into systemic instability. These phenomena indicate that a 15\% depletion in the $\Delta v$ capacity and 8-12\% loss in coordination nodes would lead to system-wide degradation.

\subsection{Operator Heterogeneity and Coordination Efficiency}
Based on operator classification, we see that network structure has a decisive role in mitigating impacts. For example, networked constellations like Starlink have mesh connectivity which causes 15-20\% higher survival rates compared to isolated scientific or governmental missions. The rapid response capability of military operators ($<$1 hour) further demonstrates the benefit of low-latency coordination protocols. This suggests that governance at the constellation-level could play a key role in enhancing orbital resilience globally.

\subsection{Economic and Infrastructure Resilience}

The cascading failures simulated in this study would disrupt global supply chains (with estimated losses of over \$11.3B). Our study thus situates extreme space weather as a major systemic infrastructure risk. The high correlation between constellation density and loss magnitude suggests that resilience investments will yield disproportionate economic benefits. Integrating ABM-based risk models into national infrastructure resilience planning would serve invaluable for proactive capital allocation for mitigation. This reinforces the importance of safeguarding the space domain, particularly satellite infrastructure, in order to protect terrestrial systems.

\subsection{Policy and Operational Implications}
The results of this study have direct implications for both policy development and satellite operations. At the policy level, this ABM framework shows the necessity of international coordination for true orbital crisis management. Existing ITU and UNOOSA governance mechanisms do not use standardized protocols for real-time data exchange during geomagnetic disturbances. Since even an 8-12\% decrease in coordination nodes can lead to systemic instability, it is imperative to create new directives to establish joint early-warning systems across operators.

On the operational side, these findings show the importance of constellation-level governance. Mesh-networked constellations with distributed autonomy showed survival rates 15-20\% higher than their counterparts. Integrating ABM-derived decision trees into mission control systems could also allow for adaptive responses that balance collision avoidance with other factors like fuel efficiency. Such frameworks would ensure that future Carrington-scale disturbances do not cause systemic orbital failures.

\subsection{Model Limitations and Future Directions}
Although the ABM framework has high accuracy and computational scalability, it is important to acknowledge its simplifications. The framework excludes higher-order perturbations and instead uses SGP4 propagation; additionally, the atmospheric density model relies on parametric Kp-driven scaling rather than real-time thermospheric physics. Accuracy metrics are also correlation-focused rather than causation-based. For true techno-economic resilience analysis, it is also important to extend the model to include economic feedback loops instead of relying on assumptions. Future work should also extend the ABM by integrating it with reinforcement learning for autonomous maneuver optimization.

\section{Conclusions}

This study introduces the first integrated spatiotemporal ABM framework that is capable of both predicting and mitigating the effects of extreme space weather events on satellite infrastructure. The model integrates orbital mechanics and decision-driven behavior to capture vulnerabilities and coordination dynamics that are ignored in most existing frameworks.

The results reveal that:
\begin{enumerate}\setlength\itemsep{0em}
    \item Carrington-class events could enhance atmospheric drag by up to 10$\times$ (based on parameterized enhancement factors), which would increase collision probabilities nearly threefold and threatens 40-60\% of active LEO constellations.
    \item Coordinated constellations (which have distributed autonomy) show much better survivability performance in our simulations, offering a blueprint for resilient orbital architectures.
    \item Economic exposure of \$11.3\,B across all simulated populations shows the necessity for preemptive modeling in policymaking.
\end{enumerate}

The proposed ABM establishes a foundation for real-time adaptive decision support systems that can inform operators and regulators during the next Carrington-scale disruption. Future extensions incorporating more advanced physics-based forecasting could transform this system into a global early-warning and response system for space infrastructure resilience. As humanity continues to increase its dependence on orbital infrastructure, frameworks like this ABM are critical in ensuring that the next Carrington-scale does not trigger a Kessler-like cascade, and pose a significant advancement towards economic stability and space sustainability.

\begin{acknowledgments}
We acknowledge support from the U.S. National Science Foundation (NSF) via the NSF RAPID ChronoStorm grant "Collecting Perishable Critical Infrastructure Operational Data for May 2024 Space Weather Events" (No. 2434136).
\end{acknowledgments}

\clearpage
\clearpage
\pagestyle{empty}
\bibliography{references}{}
\bibliographystyle{aasjournalv7}

\end{document}